\documentclass{ieeetj}
\usepackage{amsmath, amsfonts}
\usepackage{graphicx, color}
\usepackage[colorlinks=true,urlcolor=blue]{hyperref}
\usepackage{multirow, makecell}
\usepackage{ragged2e}

\graphicspath{{./img/}}

\usepackage[backend=biber,hyperref,doi=false,isbn=false,style=ieee]{biblatex}
\defbibheading{bibliography}[]{}

\AtBeginDocument{
	\definecolor{tmlcncolor}{cmyk}{0.93,0.59,0.15,0.02}
	\definecolor{NavyBlue}{RGB}{0,86,125}
	\definecolor{tablecolor}{RGB}{51,133,255}
}

\def\OJlogo{\vspace{-4pt}$<$Society logo(s) and publication title will appear here.$>$}
\def\seclogo{\vspace{10pt}$<$Society logo(s) and publication title will appear here.$>$}
\def\authorrefmark#1{\ensuremath{^{\textbf{#1}}}}

\title{An Open Source Realtime GPU Beamformer for Row-Column and
       Top Orthogonal to Bottom Electrode (TOBE) Arrays}

\author{Randy Palamar\authorrefmark{1}, Graduate Student Member, IEEE,
        Darren Dahunsi\authorrefmark{1}, Graduate Student Member, IEEE,
        Tyler Henry\authorrefmark{1}, Graduate Student Member, IEEE,
        Mohammad Rahim Sobhani\authorrefmark{1}, Member, IEEE,
        and Roger Zemp\authorrefmark{1}, Member, IEEE}
\affil{Department of Electrical and Computer Engineering, University of Alberta, Edmonton, AB T6G 2R3, Canada}
\corresp{Corresponding authors: Randy Palamar (email: palamar@ualberta.ca),
         Roger Zemp (email: rzemp@ualberta.ca).}
\authornote{Funding for this work was provided by the National
Institutes of Health (1R21HL161626-01 and EITTSCA R21EYO33078),
Alberta Innovates (AICE 202102269, CASBE 212200391 and LevMax
232403439), MITACS (IT46044 and IT41795), CliniSonix Inc., NSERC
(2025-05274), the Alberta Cancer Foundation and the Mary Johnston
Family Melanoma Grant (ACF JFMRP 27587), the Government of Alberta
Cancer Research for Screening and Prevention Fund (CRSPPF 017061),
an Innovation Catalyst Grant to MRS, and INOVAIT (2023-6359). We
are grateful to the nanoFAB staff at the University of Alberta for
facilitating array fabrication.}

\receiveddate{XX Month, XXXX}
\reviseddate{XX Month, XXXX}
\accepteddate{XX Month, XXXX}
\publisheddate{XX Month, XXXX}
\currentdate{XX Month, XXXX}
\doiinfo{XXXX.2022.1234567}

\begin{document}

\begin{abstract}

Research ultrasound platforms have enabled many next-generation
imaging sequences but have lacked realtime navigation capabilities
for emerging 2D arrays such as row-column arrays (RCAs). We
present an open-source, GPU-accelerated reconstruction and
rendering software suite integrated with a programmable ultrasound
platform and novel electrostrictive
Top-Orthogonal-to-Bottom-Electrode (TOBE) arrays. The system
supports advanced real-time modes, including cross-plane
aperture-encoded synthetic-aperture imaging and aperture-encoded
volumetric scanning. TOBE-enabled methods demonstrate improved
image quality and expanded field of view compared with
conventional RCA techniques. The software implements beamforming
and rendering kernels using OpenGL compute shaders and is designed
for maximum data throughput helping to minimize stalls and
latency. Accompanying sample datasets and example scripts for
offline reconstruction are provided to facilitate external
testing.

\end{abstract}


\maketitle

\section{INTRODUCTION}

\IEEEPARstart{T}{he} field of medical ultrasound has seen rapid
developments in 2D and 3D beamforming techniques. Recently
developed Top-Orthogonal-to-Bottom-Electrode (TOBE) arrays
\cite{sampaleanu2014top}, including those based on the
electrostrictive relaxor PMN-PT, are sensitive to applied DC bias
voltages which can be used to electronically control the polarity
of elements \cite{latham18simu, sobhani2021bias}. In turn, this
has enabled spatial aperture encoding techniques
\cite{chiao1997sparse} allowing for the development of the Fast
Orthogonal Row Column Electronic Scanning (FORCES) and Ultrafast
FORCES (uFORCES) methods \cite{cc17_fast, MRS22_uFORCES}. In
comparison to the Virtual Line Source (VLS) \cite{vls2002,
rasmussen15three} and Tilted Plane Wave (TPW) \cite{flesch17}
methods achievable with traditional row-column arrays (RCAs) the
FORCES method has been shown to produce B-Scan images with
visibility beyond the shadow of the aperture and with enhanced
contrast and resolution \cite{palamar25tobe}. It is important to
note that both the VLS and TPW methods can be performed with a
TOBE array by applying a constant bias voltage across the
elements.  However, the bias sensitivity of TOBE arrays has
additionally enabled the development of the Hadamard Encoded Row
Column Ultrasonic Expansive Scanning (HERCULES) method
\cite{dahunsi25hercules}, which allows readout from every element
of the TOBE array instead of readout of only the rows or columns.
Like VLS and TPW the HERCULES method produced a 3D dataset with
similar contrast and resolution, however like FORCES, was able to
image beyond the shadow of the aperture. Both FORCES and HERCULES
require an additional software based decoding step in addition to
typical Delay and Sum (DAS) beamforming.

To advance the FORCES and HERCULES techniques towards adoption for
diagnostic purposes realtime beamforming is needed. Additionally,
as new TOBE array designs are being actively developed, we require
full customization of all array parameters. Existing openly
available solutions, such as RTBF \cite{hyun2019open}, are overly
focused on MATLAB, require NVIDIA GPUs, and are not generally
designed with realtime display and user input in mind. RTBF for
example requires the user to read back beamformed data to the CPU
so that MATLAB can send it back to the GPU to be displayed on the
screen. Methods with such limitations seem common in the
literature \cite{khan2020realtime, boonleelakul16ultrasound}. A
more performance oriented solution presented in
\cite{praesius25realtime, jorgensen23fastvolume} is not publicly
available and relies on an approximation that can only be made
when the receive aperture is composed of long line elements. The
HERCULES receive aperture resolves to a fully populated 2D grid
and therefore the previous approaches in \cite{praesius25realtime,
jorgensen23fastvolume} do not apply. Other solutions exist in the
literature but are also not freely available or only support 2D
imaging \cite{yiu2019live, nahas2025bedside}.

In this work we implement a cross-platform and multi-architecture
realtime beamforming application with live updates and controls.
We use it to demonstrate the first realtime realizations of the
FORCES and HERCULES imaging methods suitable for live imaging. The
software is configurable programmatically and via user inputs. In
addition to traditional B-Scan views the software can be
configured to display live 2D or 3D cross plane views and 3D
fly-through views. The software is fully open source and can be
operated entirely independently from MATLAB. OpenGL based and CUDA
based implementations of different processing stages are
discussed. We present performance metrics from both a desktop
system with ultrasound acquisition hardware and from a mobile
ultrabook performing offline data processing. Furthermore, it is
imaging system agnostic, with all communication occurring through
an application binary interface (ABI). A helper library which
implements a C application programming interface (API) is
provided, and we use it to interface with a Verasonics (Kirkland,
WA, USA) Vantage-256 Research Ultrasound Platform to perform
realtime imaging.

We present the article as follows: first we discuss the
capabilities of the implemented beamformer and give some examples
of usage, then we provide some theoretical context needed to
describe the optimizations performed. A brief introduction to the
hardware utilized for testing is provided, and a description of
the methods used to communicate with the application and the
available data processing methods are introduced. We conclude the
article with a detailed description of a number of software
optimizations we performed, and provide a direction for future
works.

\section{RESULTS \& DISCUSSION}

We implemented both the beamformer and a library used for
interfacing with it in the C programming language. GPU code was
implemented in the OpenGL Shading Language (GLSL) \cite{glslspec}.
C was chosen due to its simplicity and its ability to access
native CPU features and low level operating system functions with
little overhead. For the interface library, C's status as a
{\em{lingua franca}} \cite{cstandardcharter} means that bindings
for other languages such as Python or MATLAB can be written
without significant burden; some simple MATLAB bindings to this
library are provided with the software. We utilized OpenGL
\cite{openglspec}, for all rendering and beamforming, and Raylib
\cite{raylib}, for simple shape drawing (used in the user
interface). NVIDIA's CUDA \cite{cuda2025} is supported at runtime
providing access to a Hilbert transform and an alternate version
of the decoding stage. However, unlike MATLAB's GPU API
\cite{matlabGPU}, the majority of our GPU code does not require an
NVIDIA GPU to run. No other libraries were utilized and any needed
functionality was implemented from scratch. Additionally we did
not utilize any prepackaged build system for the project. Instead
all required functionality is provided in the included build tool
(also written in C). The only requirement for building the
application is a modern C compiler supporting C11 with some
commonly available extensions and intrinsics.

\subsection{User Interface (UI)}

\begin{figure}
	\centering
	\includegraphics[width=\columnwidth]{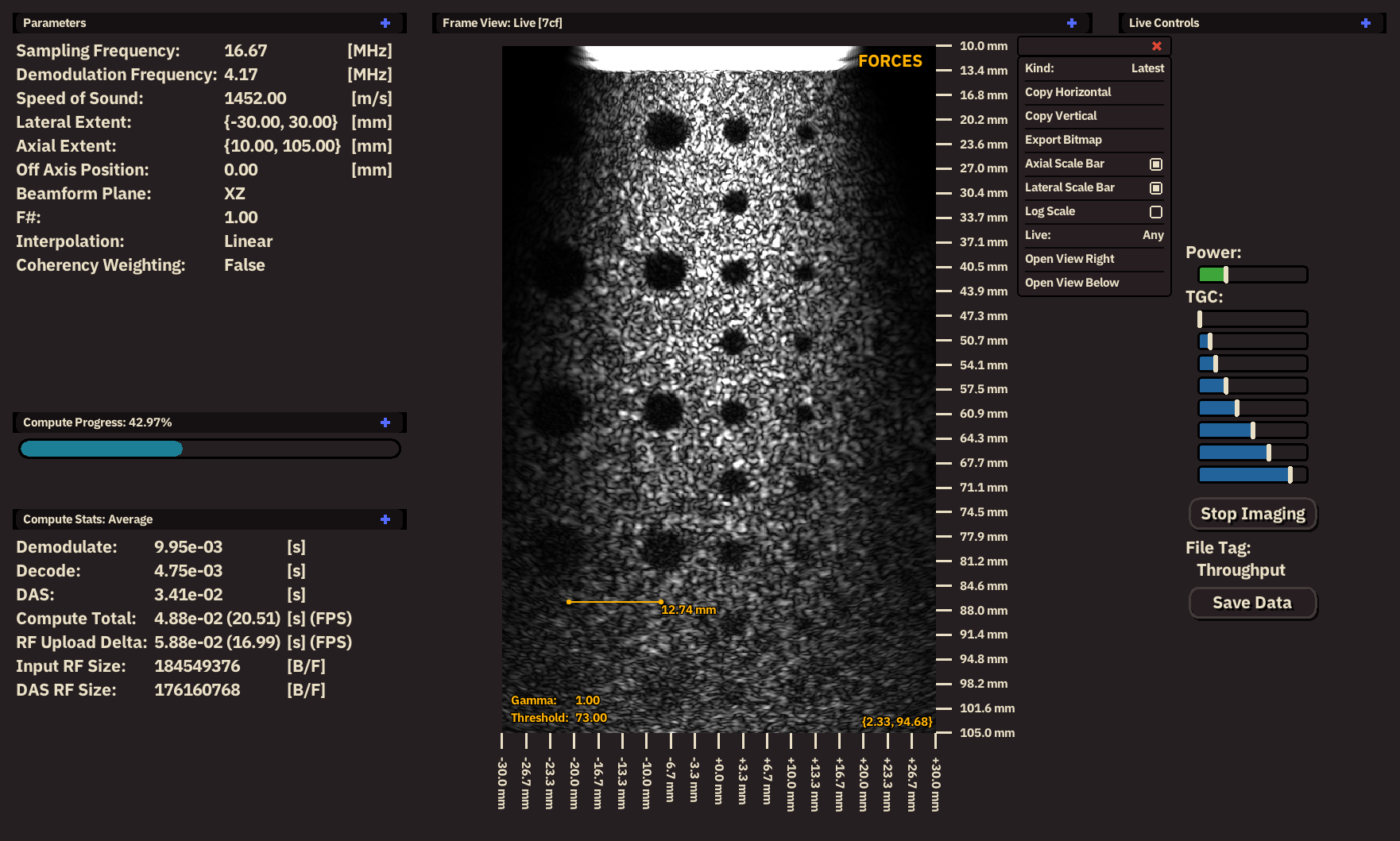}
	\caption{The default application view when performing realtime imaging.
	         Most visible parameters are adjustable at runtime but the live controls
	         on the right involve additional support from the imaging system. Here
	         the imaging sequence utilized a chirp excitation so the top of the
	         beamformed image is part of the deadzone. This is shown for demonstration
	         only, typically we would not beamform this region. The view can be changed
	         from the shown B-Scan mode to other modes such as a 3D Cross-Plane view by
	         adjusting the `Kind' field in the drop down.}
	\label{fig:basic_ui}
\end{figure}

\begin{figure}
	\centering
	\includegraphics[width=0.6\columnwidth]{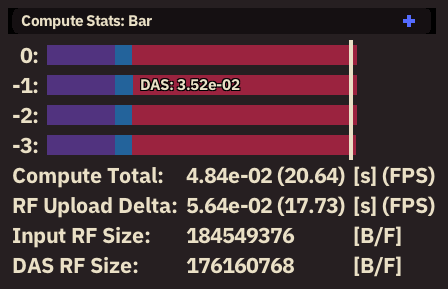}
	\caption{Compute stats view showing the proportion of time taken by each
	         shader during beamforming. When any particular portion is hovered
	         a text is shown with the shader's name and the amount of time taken
	         in that frame. The white bar on the right represents the 32-frame average
	         total beamforming time. The input RF size is the size of data uploaded to
	         the GPU, and the DAS RF size is the size after data is converted to (complex)
	         floating point values and (optionally) decimated.}
	\label{fig:basic_ui_bar}
\end{figure}

The application provides a user interface (UI) for adjusting many
imaging parameters at runtime. Figure \ref{fig:basic_ui} shows an
example of the application. On the left of the application we list
all parameters used for beamforming. Apart from the sampling
frequency and demodulation frequency, which are purely
informational, all parameters may be adjusted during live imaging.
The floating window in the top right is opened by clicking on the
`$+$'. This allows the user to make a fixed copy of the currently
visible image, which is useful for comparing different modes or
parameters, change the display to a log scale, and more. The live
display can be scrolled to adjust the power threshold
(brightness), when displaying on a power scale, or the dynamic
range when displaying on a log scale. The display is always aspect
ratio correct. By left clicking on the image the ruler, visible in
the figure is opened. This can be used to measure features or
distances. The text in the bottom right indicates the current
position of the mouse cursor converted to millimeters. On the far
right are live imaging parameters which can be used to adjust the
imaging system's parameters at runtime. This may also be used to
save data after a region of interest has been located. Finally the
bottom left of the UI displays performance metrics related to the
beamforming. The times shown are 32-frame averages to help smooth
out inconsistencies between frames. We list both the total compute
time and the time between RF data uploads so that we know which
part of the system is limiting performance. In most cases we are
limited by the data upload rate, but in cases where we are limited
by computation time, the two values will converge as we do not
allow unprocessed data to be overwritten. The view can be switched
to a bar graph showing the shader time breakdown for the last 4
frames giving a visible representation of the proportion of frame
time taken by each processing stage. This is shown in Figure
\ref{fig:basic_ui_bar}. Since the UI operates entirely
independently, slowdowns during beamforming do not cause any
noticeable delays in the user interface.

\subsection{Capabilities}

The implemented beamformer supports both IQ beamforming and direct
RF beamforming. It supports data formatted as 16-bit integers,
16-bit complex integers, 32-bit floating point values, or 32-bit
complex floating point values. Both Linux and Windows are
supported, along with both AMD64 CPUs and ARM64 CPUs. Furthermore,
in addition to supporting the imaging methods developed by our
lab, it supports conventional RCA imaging methods (VLS
\cite{vls2002, rasmussen15three}, TPW \cite{flesch17}) and 1D
array imaging (Walking Aperture, SA, Flash). These specifications
are listed in Table \ref{tab:capabilities}.

\begin{table}
	\centering
	\caption{Beamformer Specifications \& Features}\label{tab:capabilities}
	\begin{tabular}{p{0.2\columnwidth}>{\RaggedRight\arraybackslash}p{0.72\columnwidth}}
		\textbf{Methods}       & (u)FORCES, (u)HERCULES, VLS, TPW / OPW, Flash, Walking-FORCES,
		                         OPTIMUS, HERO-PA \\
		\textbf{Data Types}    & Int16, Int16-Complex, Float32, Float32-Complex                       \\
		\textbf{Output}        & 2D or 3D Float32/Float32-Complex Data                                \\
		\textbf{Interpolation} & Nearest, Linear, Cubic (Hermite)                                     \\
		\textbf{Data Size}     & 2GB in a single call                                                 \\
		\textbf{OS}            & Windows, Linux                                                       \\
		\textbf{Hardware}      & AMD64/ARM64 CPUs, any GPU supporting OpenGL 4.5                      \\
		\textbf{Arrays}        & TOBE, RCA, Linear                                                    \\
		\textbf{Views}         & B-Scan, Cross-Plane B-Scan, 3D-Sweep, 3D Cross-Plane                 \\
		\textbf{Features}      & Adjustable FOV; Intensity Transformations (Log, Power); Arbitrary
		                         Waveform Matched Filtering; Multi-Array Imaging; Dynamic Receive Apodization;
		                         Coherency Weighting; GPU Timings; TGC Control; Transmit Power Control
	\end{tabular}
\end{table}

\subsection{Cross Plane Navigation}

\begin{figure}
	\begin{minipage}[c]{0.49\columnwidth}
		\centering
		\includegraphics[width=\textwidth]{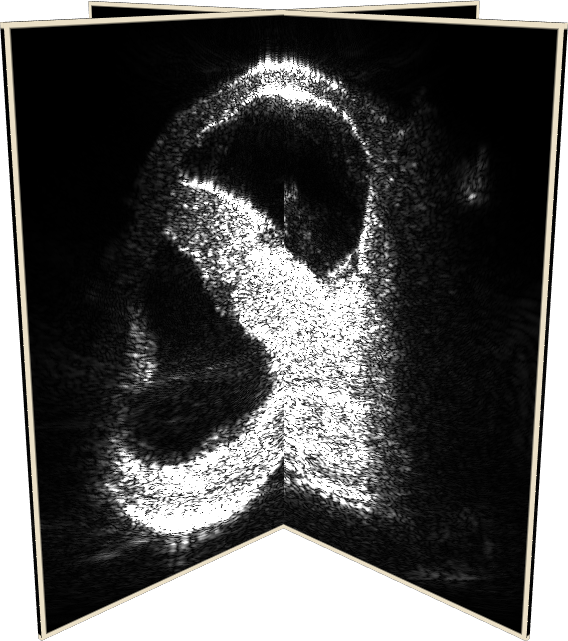}

		(a)
	\end{minipage}\hfill
	\begin{minipage}[c]{0.49\columnwidth}
		\centering
		\includegraphics[width=\textwidth]{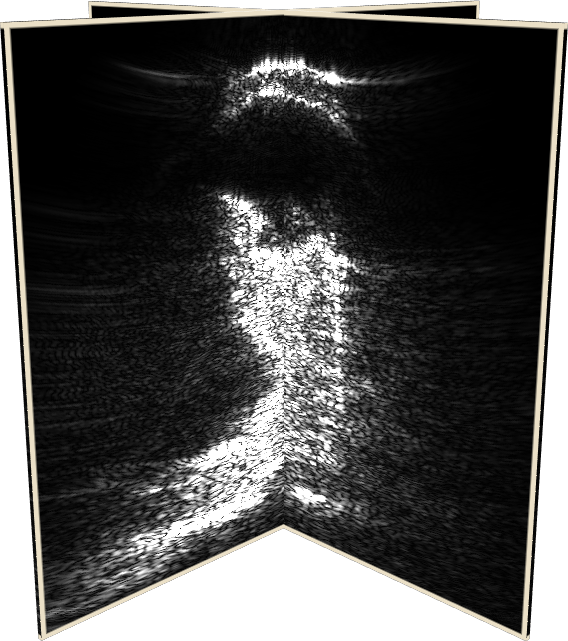}

		(b)
	\end{minipage}\hfill
	\caption{Example of navigation using (a) Cross-Plane FORCES vs (b) Cross-Plane VLS.
	         FORCES ability to see beyond the shadow of the aperture and its enhanced contrast
	         aim to improve the operator's ability to navigate in-vivo. This image of a water
	         immersed heart phantom starts at a depth of 2.5cm and ends at a depth of 13cm with
	         each plane having a width of 12cm. The view shown utilizes a perspective camera and
	         therefore a scale bar would be meaningless and is not provided. We compounded 128 low
	         resolution images for each visible plane.}
	\label{fig:xplanes}
\end{figure}

While many modern clinical ultrasound systems contain 3D
navigation capabilities, limitations such as low-framerate and low
image quality can lead to them being unused in practice
\cite{hung20073d, huang2017review, poon20193d,
rutledge2020commercial}. In Figure \ref{fig:xplanes} we
demonstrate a 3D cross-plane view. This is an alternate view
provided by our software which is accessed by selecting the
appropriate menu item in the drop down. We propose that by
visually highlighting the full 3D context this view is better
suited for navigation than individual B-Scan planes. While
conventional RCA methods are able to create such planes, provided
they remain beneath the shadow of the probe, FORCES' unique
ability to focus a single elevational plane in transmit makes it
better suited for such a use case. Additionally FORCES is able to
image beyond the aperture's shadow extending the view rectangle
into a view frustum. To demonstrate the realtime capabilities of
our Cross-Plane FORCES method we utilized a beating heart phantom
(Shelly Medical Imaging Solutions, London ON, Canada). A video
with the heart beating at 1Hz is provided in Supplementary Video
1. The location of each imaging plane may be adjusted during
operation by selecting and dragging to the desired position as
shown in Supplementary Video 2. Previous work \cite{palamar25tobe}
has demonstrated that FORCES can be electronically walked/scanned
to map out a high resolution volume. Supplementary Video 3
demonstrates this capability by updating one of the imaging planes
between each acquisition during a Cross-Plane imaging session.

Although FORCES can generate high-resolution volumes by scanning,
it requires many emissions (e.g., $N\times128$ for $N$ slices) and
is therefore best suited for producing B-Scan images for 2D
navigation. We propose using FORCES to locate a region of interest
before volumetric acquisition: once the target region is
identified, a volumetric method such as HERCULES, VLS, or TPW can
collect the dataset. Because these volumetric methods rely on
unfocused transmits, they do not achieve the same contrast as
FORCES \cite{palamar25tobe, dahunsi25hercules}, making small
vesicles and other low-contrast features difficult to detect.
Regions containing such features are commonly selected for
Ultrasound Localization Microscopy (ULM), Speckle Decorrelation
Imaging, and Vector Flow Imaging \cite{nahas2025bedside,
ketterling2017high, cohen2024multidirectional, zhou20183,
wang2024ultrasound, shin2025high, lowerison2025comparison,
wu20253d}. However, without extensive post-processing it can be
hard to verify that the correct region was targeted. We propose
that Cross-Plane FORCES will enable high-quality cross-plane
navigation that is not achievable with conventional RCAs.

\subsection{Performance}

\begin{table}
	\centering
	\caption{Performance Comparison Parameters}
	\label{tab:perf_comp_params}
	\begin{tabular}{lc}
		Input Samples        & 2816 \\
		Channels             & 128  \\
		Emissions            & 128  \\
		Total Samples        & 92M  \\
		Interpolation        & Cubic (Hermite) \\
		B-Scan Filter Length & 166  \\
		B-Scan Points        & 1024$\times$1024 \\
		Volume Filter Length & 36   \\
		Volume Points        & 256$\times$256$\times$256\\
	\end{tabular}
\end{table}

\begin{table}
	\centering
	\caption{Performance Comparison}
	\label{tab:perf_comp}
	\begin{tabular}{lcccc}
		Method                           & \makecell{NVIDIA \\ RTX 4090} & \makecell{AMD \\ RX 9070 XT} & \makecell{Qualcomm \\ Adreno X1} \\
		\hline
		\hline
		\makecell[l]{FORCES   \\ B-Scan} & 61.916                        & 29.651                       & 681.983  \\\hline
		\makecell[l]{HERCULES \\ B-Scan} & 66.943                        & 42.059                       & 700.653  \\\hline
		\makecell[l]{TPW      \\ B-Scan} & 64.078                        & 34.640                       & 640.375  \\\hline
		\makecell[l]{HERCULES \\ Volume} & 62.441                        & 59.701                       & 1047.424 \\\hline
		\makecell[l]{VLS-128  \\ Volume} & 60.080                        & 59.570                       & 1082.909 \\\hline
	\end{tabular}
	\vspace{0.25em}

	* All values in nanoseconds/point
\end{table}

Performance of the application was tested with three different
GPUs: an NVIDIA RTX 4090, an AMD RX 9070XT, and a Qualcomm Adreno
X1 (San Diego, CA, USA). We measure the performance for B-Scans
using FORCES, HERCULES, and TPW, and volumes using HERCULES and
VLS. All methods utilized 128 emissions and received on 128
channels. The B-Scans utilized a Matched Chirp Filter with 166
taps and the volumes used a simple low pass filter with 36 taps.
1M points were beamformed for the B-Scans, and 16M were beamformed
for the volumes. All parameters are listed in Table
\ref{tab:perf_comp_params}. Timings were measured using
asynchronous GPU side timers as they are the only way to ensure
that we are actually measuring the time taken by the GPU.

Timings for each case and each GPU are provided in Table
\ref{tab:perf_comp}. The highest performance was achieved with the
AMD GPU. For the FORCES method 29.6 ns/point was reached which is
high enough to obtain 60 frames per second (FPS) with $\sim$500K
points per image. As will be discussed in Section
\ref{sec:optimizations}.\ref{sec:vantage}, this exceeds our
current hardware's capabilities, which limit us to just 24 FPS
during live imaging. For the volumetric methods we achieved 59.7
ns/point which is only able to serve $\sim$ 1 Volume per second at
16M points. While this is significantly lower than the numbers
reported in \cite{praesius25realtime}, we are processing 4.2x the
data, beamforming 4x the points, and using a more expensive
interpolation method. Our method also does not rely on
pre-acquiring many volumes worth of data, which limited the rate
at which they could display those volumes to just 10FPS. As
mentioned above, the algorithm presented in
\cite{praesius25realtime, jorgensen23fastvolume} does not apply to
the HERCULES method so we are not able to benefit from its
advantages. The performance on the Qualcomm GPU is roughly 20x
slower than the AMD GPU. This is partially accounted for by a 10x
lower floating point operations per second (FLOPS), with the rest
likely accounted for by the limited number of available registers
and significantly lower cache size. On paper the RTX4090 should be
roughly 40\% faster than the AMD GPU but in our tests lagged
behind by $\sim$2x for B-Scan images, and a small amount for the
volumes. This is likely due to the software seeing significantly
more testing and optimization work on a system with an AMD GPU. It
indicates that the while the underlying architecture of the two
GPUs is similar it is not identical. Furthermore, AMD's Linux
driver is likely more robust due to the nature of its open source
development which has allowed for many different parties to
contribute fixes and optimizations for a wide range of use cases.
With further testing on NVIDIA GPUs a set of codepaths better
suited to NVIDIA's architecture can be developed to reflect their
on paper performance advantages.

Another important aspect of the program's performance is the
achievable data transfer rate. A highly optimized overlapped
transfer design was implemented which can achieve sustained
transfer rates of up to 18GB/s into a buffer on the GPU. This
exceeds what is possible with PCIe3.0 x16 (15.7GB/s) which is the
maximum available on current research ultrasound platforms. Having
such a high data rate can aid in performing ultrafast volume
acquisitions where the desired Pulse Repetition Frequency (PRF)
can reach 20kHz \cite{flesch17, sauvage2018large}. The beamformer
application supports saving these acquisitions for later offline
processing.

\section{EXAMPLES}

We provide an example script for offline reconstruction in
Supplementary File 1. The script requires a number of additional
files provided by the beamformer. Prebuilt versions of the
beamformer suitable for running on Windows and packaged with all
required files are available online
(\href{https://github.com/UAlberta-Zemp-Lab/ornot/releases}
{github.com/UAlberta-Zemp-Lab/ornot/releases}). The script expects
a running version of the beamformer on the same computer or an
error will be produced. Additionally, the script is written to
accept a data format described in the online release, however the
script is easily modifiable to directly accept an arbitrary data
array, for example a RcvBuffer from a Vantage system. A test
dataset is povided online.

Reconstruction during live imaging uses a more advanced API
designed to allow a higher degree of control by the user. An
example in C is provided in the beamformer's tests folder
(\href{https://github.com/rnpnr/ogl_beamforming/tests/throughput.c}
{github.com/rnpnr/ogl\_beamforming/tests/throughput.c}).
Supplementary File 2  provides an example of the setup and usage
for performing live reconstruction. The `oglBeamform' function is
entered via an anonymous function handle called during a
ReturnToMatlab Vantage event.

\section{THEORY}\label{sec:theory}

Here we provide a brief explanation of two key aspects of how the
ultrasound data is handled. First we describe Synthetic Transmit
Aperture imaging and how it relates to the FORCES and HERCULES
methods. Then we discuss sampling and interpolation, in particular
the aspects which can be utilized to reduce processing
requirements.

\subsection{Synthetic Aperture Imaging}\label{sec:sta}

Synthetic Aperture (SA) imaging is a well known technique used in
practice for many decades \cite{burckhardt1974experimental,
corl1980real, jensen2006synthetic}. It consists of firing a
sequence of transmits each using a small subset of the aperture,
typically just a few elements, and then receiving on the entire
array. Each of these individual transmits may be beamformed into a
Low Resolution (LR) image (or volume) by applying a delay-and-sum
(DAS) to the receive dataset. All LR images are summed together to
form a high-resolution (HR) image. This is described by the
following equation:

\begin{equation}\label{eq:das}
	\text{HR}(x,z) = \sum_{i = 1}^{N_\text{tx}} a_i^{\text{tx}}(x,z)
	             \sum_{j = 1}^{N_\text{rx}} a_j^{\text{rx}}(x,z) r_{i,j}\left(\underset{i,j}{\text{ToF}}(x,z)\right)
\end{equation}

here $a^\text{tx}$ and $a^\text{rx}$ represent apodization applied
for each transmit and receive, $r_{i,j}(t)$ is the receive data
set for transmit $i$ and receiver $j$, and $\text{ToF}_{i,j}(x,z)$
is the time of flight to a particular image point. A standard, but
not optimal \cite{schiffner2023frequencyfnumber}, definition for
the receive apodization based on maintaining a constant F-Number
is:

\begin{equation}\label{eq:receive_apodization}
	a_j^{\text{rx}}(x,z) = W\left(F_\#\frac{|x - x_j|}{|z - z_j|}\right)
\end{equation}

where $(x_j, z_j)$ is the spatial location of the $j$-th receiver
(typically $z_j = 0$) and $W$ is a windowing function which we
chose to be a Hanning window:

\begin{equation}\label{eq:window}
	W(u) = \begin{cases}
		\cos^2\left(\pi u\right) & |u| < 0.5 \\
		0                        & \text{otherwise}
	\end{cases}
\end{equation}

Transmit apodization $a^\text{tx}$ is typically only applied for
focused transmits and is chosen based on the focal depth.
Generally it is applied in hardware but the coefficients are
chosen using the same equation above.

While the FORCES method applies a spatial Hadamard encoding across
the array, after the data is decoded the final dataset resolves to
one where it was as if the first transmit occurred on the first
element, the second transmit occurred on the second element, and
so on. In fact, the only difference is that for each transmit the
entire aperture was utilized instead of just a single element.
This greatly increases the amount of transmit energy which in turn
increases the Signal to Noise Ratio (SNR) of the image by a factor
$\sqrt{N_\text{tx}}$ \cite{misaridis2002space} and increases the
attainable imaging depth. The HERCULES method is similar but
instead of applying the encoding to the transmit aperture it is
applied to the receive aperture. While physically the receive
aperture is composed of long elements, after decoding, the
recovered dataset is one where it was as if we received from every
element of a 2D transducer array \cite{dahunsi25hercules}. This is
only possible due to the bias sensitivity of the electrostrictive
relaxor material used in the manufacturing of TOBE arrays. In
particular, the phase inversion that the material imparts on the
signal \cite{latham18simu} enables the simultaneous
transmit-receive focusing of FORCES and the recovery of a 2D
receive aperture with HERCULES. Fundamentally there is no
difference between the final dataset after decoding and a linear
array SA dataset for FORCES, or a 2D matrix array emitting plane
or diverging waves for HERCULES. FORCES SNR improvement is
entirely due to transmit focusing and the use of the entire
aperture during emission.

\subsection{Interpolation \& Sampling}\label{sec:sampling}

While the received ultrasound signal is continuous it must be
sampled if we wish to process it with a digital system. It is well
established \cite{shannon1949communication} that in order to fully
reconstruct an arbitrary signal it must be sampled at $2f_c$ where
$f_c$ is the maximum frequency component of the signal. However,
ultrasound signals are typically bandlimited. We can therefore
utilize quadrature sampling to reduce the sampling requirements
and equivalently reduce the amount of data which must be processed
\cite{powers1980ultrasound}. This simply requires that two samples
be taken exactly $90^\circ$ out of phase from one another. To
reconstruct the original signal from the received digital samples
interpolation must be used. A complete representation of the
signal at the sample times can be obtained by using sinc
interpolation \cite{oppenheim2014discretech4}; however, an
approximation to the sinc interpolator, operating on just a small
window of samples, may achieved by using a polynomial spline
\cite{unser1992polynomial,unser1993b}. The degree of polynomial
may be selected by considering that the more oversampled the
signal is the fewer points are needed to accurately represent the
original \cite{oppenheim2014discretech4}. For interpolators of
degree $n > 1$ there exists multiple sets of basis polynomials to
choose from, for example Lagrange or Hermite. A more general set
of coefficients may be chosen by considering the interpolator as a
finite-impulse-response (FIR) filter \cite{schafer1973digital}.
Here we choose to use Hermite polynomials since they ensure that
the derivative of the reconstructed signal remains smooth.

To satisfy the Nyquist rate for the bandlimited quadrature signal,
assuming a fractional bandwidth of $B$ centered about the
demodulation frequency $f_d$, we must minimally sample at a rate
of $B*f_d$. Since we want to use a simple 4-point cubic hermite
spline interpolation we may choose to oversample the signal. For
example the Vantage system offers a number of sampling modes
suitable for quadrature sampling. The NS200BW mode assumes a
signal bandwidth of 200\% ($B = 2$) centered at $f_d$, and
performs Nyquist sampling at $2Bf_d$. This samples the wave at 0,
$\frac{\pi}{2}$, $\pi$, and $\frac{3}{2}\pi$ radians. Thus the
quadrature signal is sampled at twice per period with the second
sample being inverted. Furthermore, this mode is suitable for
reconstruction of the full RF signal without modification. For
reconstruction using the baseband IQ signal this is often very
oversampled, as most probes have a bandwidth $B < 1$, and can be
decimated.

\section{HARDWARE}\label{sec:hardware}

\begin{figure}
	\includegraphics[width=0.9\columnwidth]{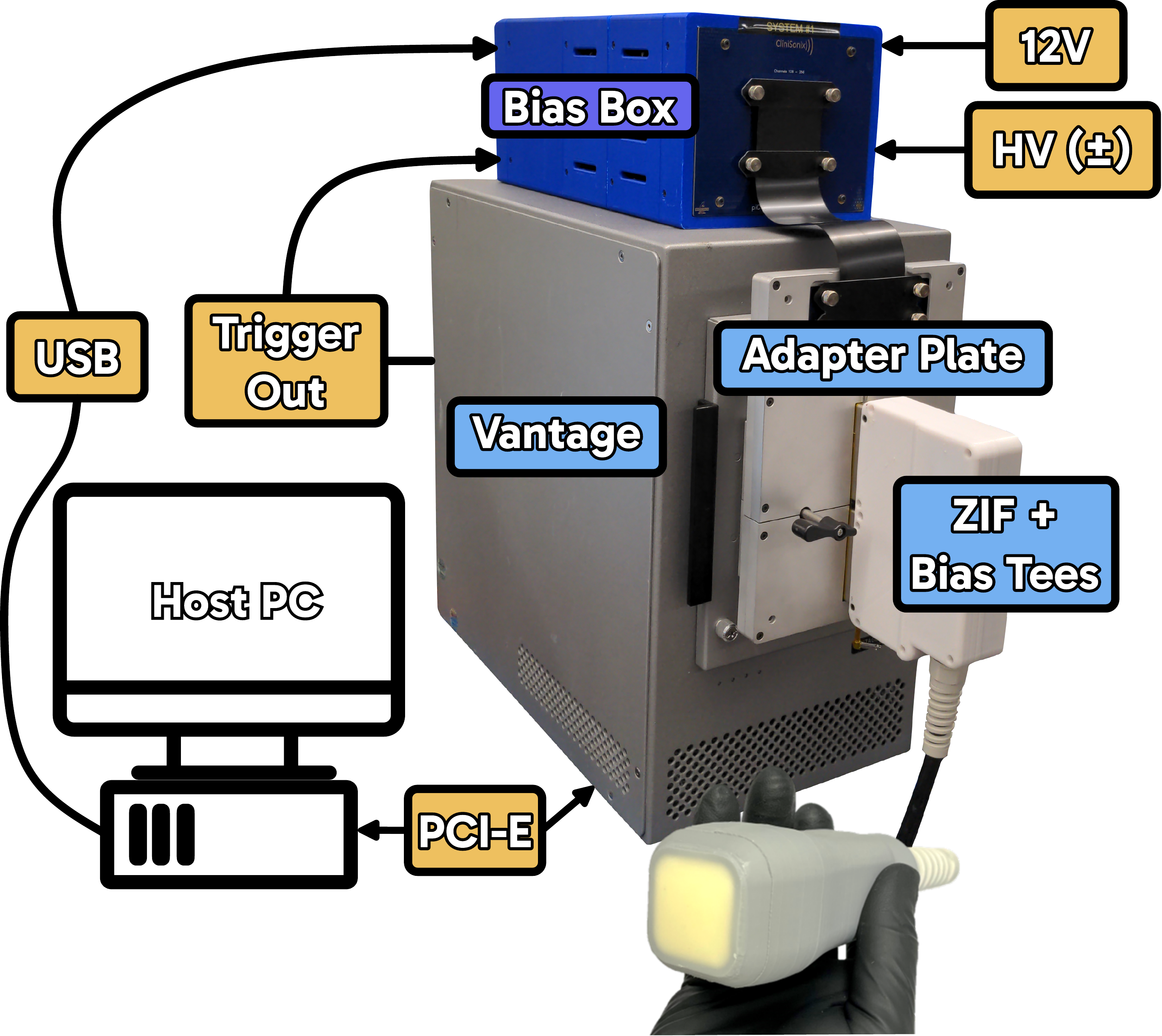}
	\caption{Live imaging system setup. In addition to the standard setup
	         we require an extra set of biasing electronics to utilize the
	         Clinisonix TOBE array shown. The biasing hardware is setup over
	         USB and controlled with a trigger out in the programmed Vantage
	         imaging sequence.}
	\label{fig:hardware}
\end{figure}

The software implemented in this study was designed to be fully
independent of a particular set of hardware, it runs on desktop
computers with discrete GPUs, and on laptops with integrated GPUs.
For the performance numbers quoted in this work however we
utilized a custom host PC with a 12-Core AMD (Santa Clara, CA,
USA) Ryzen 9 7900X3D with 96GB of DDR5 RAM. The system was
equipped with an NVIDIA (Santa Clara, CA, USA) RTX-4090 GPU with
24GB of VRAM. For the live imaging implementation we utilized a
Verasonics Vantage-256 HF Ultrasound System. Clinisonix (Edmonton,
AB) provided a 4.3MHz 128$\times$128 $\lambda$-pitch TOBE Array
with 70\% bandwidth and a set of electronics for interfacing with
the Vantage. These interfacing electronics consist of a system
providing bias voltages for the TOBE array, implemented in
\cite{ilkhechi23}, and an adapter plate for routing both bias
voltages and TX/RX channels from the Vantage to the bias-tees
located in the probe's connector. A diagram of this hardware is
shown in Figure \ref{fig:hardware}.

Data was transferred over PCIe from the Vantage system to host
system's main memory and then over PCIe from the host system to
the GPU. Due to a limitation on the number of PCIe lanes on the
host system's motherboard, transfer from the Vantage system was
limited to 4GB/s. Transfer from the host system to the GPU was not
limited and could utilize up to the maximum rate of PCIe Gen4 x16
(32GB/s). Throughput measurements were performed from the host to
GPU to find the true performance of the application when it is
fully utilized.

\begin{figure}
	\centering
	\includegraphics[width=0.6\columnwidth]{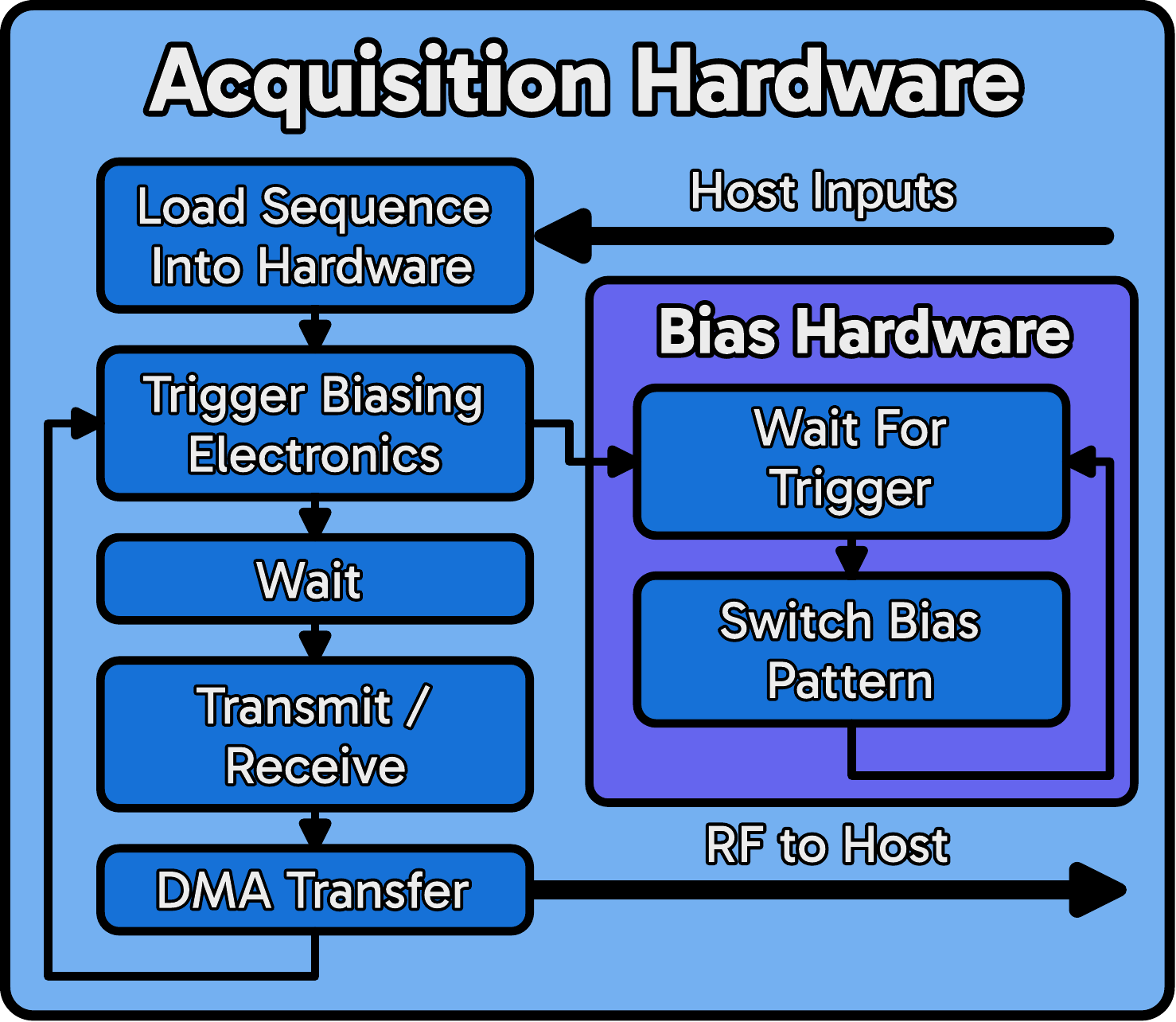}
	\caption{Asynchronous run loop of the imaging hardware. During a
	         typical run the biasing hardware is only contacted by the
	         host system at startup. Afterwards it is controlled entirely
	         by hardware triggers in the imaging sequence.}
	\label{fig:platform}
\end{figure}

During a live imaging session the host PC configures the Vantage
to perform one of the desired imaging sequences and programs the
biasing hardware with the sequence of bias voltage patterns
necessary to support the imaging method. The host PC then sets the
Vantage to run asynchronously and presents the user with the
beamformer's control interface. While running, the Vantage system
is required to send a trigger to the biasing hardware prior to
starting a transmit sequence. After triggering, the Vantage is
expected to wait for a predetermined amount of time based on the
switching characteristics of the hardware and its ability to
switch the material's polarization. For the utilized probe this
was set to 250$\mu$s allowing a 4kHZ pulse-repetition-frequency
(PRF). Once a Transmit-Receive sequence has been performed the
hardware returns the data to the host PC for processing. A visual
representation of this asynchronous loop is shown in Figure
\ref{fig:platform}.

\section{METHODS}

\begin{figure}
	\includegraphics[width=\columnwidth]{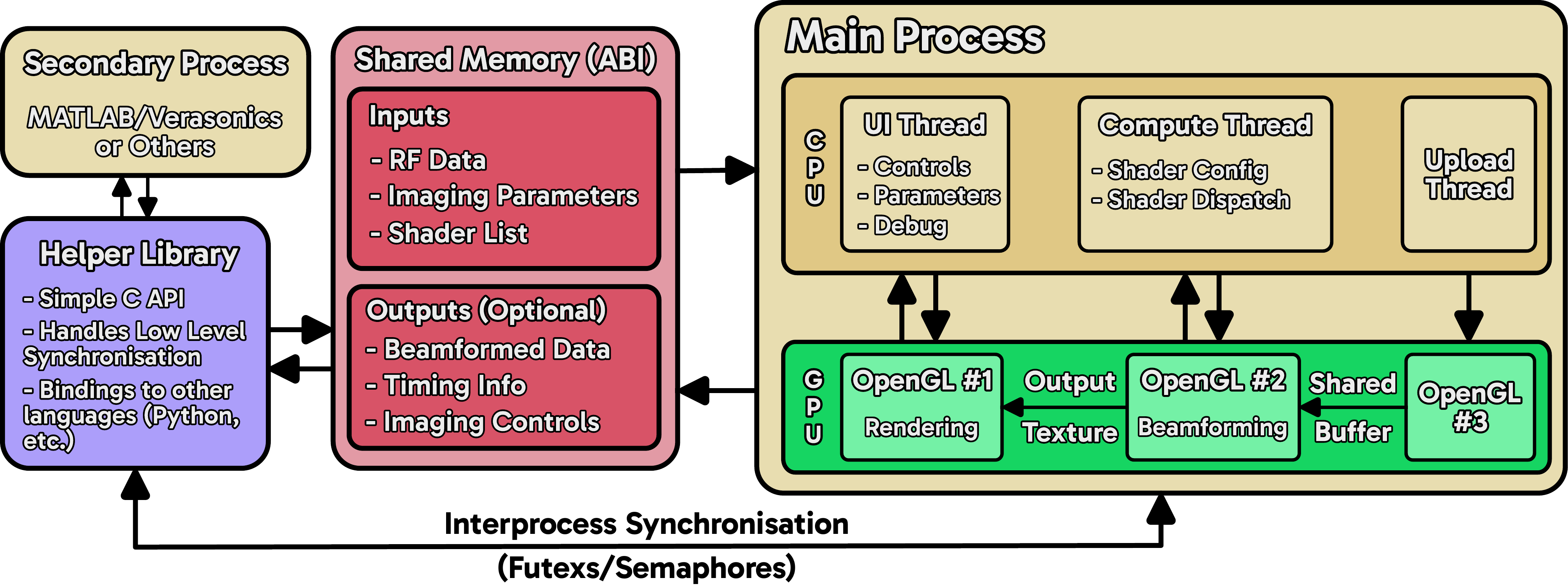}
	\caption{High level process diagram of the completed application. During
	         standard operation the main application maintains 3 separate
	         CPU threads which are configured to share GPU resources. Once an
	         image has been beamformed it will only leave the device if the
	         user has requested a readback. This helps to minimize the latency
	         between acquiring the RF data and displaying the image. A helper
	         library provides a simple interface which hides the intricacies
	         of the interprocess communication.}
	\label{fig:software}
\end{figure}

\subsection{Communication}

Communication with the beamforming application is performed
through a shared memory region. This provided the highest
performance at the cost of a higher implementation complexity and
lower runtime flexibility. Shared memory is an operating system
level feature which allows two processes to access the same region
of physical device memory at runtime. Typically modern operating
systems prevent this for security reasons \cite{intelArchManual}.
The loss of flexibility comes from the fact that both processes
must agree upon the memory region's name and size prior to opening
it. In practice this means we choose a fixed size for the region
at application compile time. Once both processes have successfully
opened the memory region, they must agree who is allowed to write
to a specific section at a particular point in time. Additionally
these writes must be atomically coherent (made visible across
threads). For register sized values this can be guaranteed by the
CPU's native atomic instructions, but for larger regions we must
employ a locking mechanism that works across processes, with both
hardware and software memory write barriers to ensure that both
the compiler and CPU perform the requested operations in the
desired order. A software lock, also known as a mutex, is a
synchronization primitive which enables two or more active
hardware threads to coherently communicate ownership of another
resource (memory, I/O device, etc.). While such a lock can be
implemented using nothing more than hardware atomic and
monitor/wait instructions, we utilize helpers provided by the
operating system so that the CPU core may be utilized by other
processes on the system. However, operating system support for
cross-process waiting can vary. Linux provides Fast Userspace
Mutexes (Futexs) which allow any memory address pointing to a
32-bit word to serve this purpose (even if the address is in a
shared memory region) but Windows provides no such mechanism
\cite{chen17comp}. Instead we must utilize Window's Semaphores
with a predetermined naming convention when running on Windows. In
order to communicate to the main application that a larger region
was modified we utilize a single flag variable with each bit
representing a different section. This variable can be easily
updated without locking by using `atomic and' and `atomic or'. For
larger regions, or sections of memory that are semantically
grouped such as beamforming parameter sets, we must first lock the
region from being modified from the other process. We then utilize
a memory copy to transfer our data of interest, employ a memory
write barrier, and release the lock. In standard operation the
main application will not be signaled until a new RF dataset is
available. Once signaled it will update the current pipeline if
the parameter set was modified (known by checking the appropriate
flag variable) and then try to beamform the data. Since the data
upload and beamforming occur in different threads, as shown in
Figure \ref{fig:software}, the primary application thread will
need to wake them the first time data is available. When the
library caller has indicated that live imaging is being performed
the upload and compute threads will attempt to reduce latency and
maximize throughput by never yielding control back to the
operating system.

\subsection{Data Processing}

\begin{figure}
	\centering
	\includegraphics[width=0.92\columnwidth]{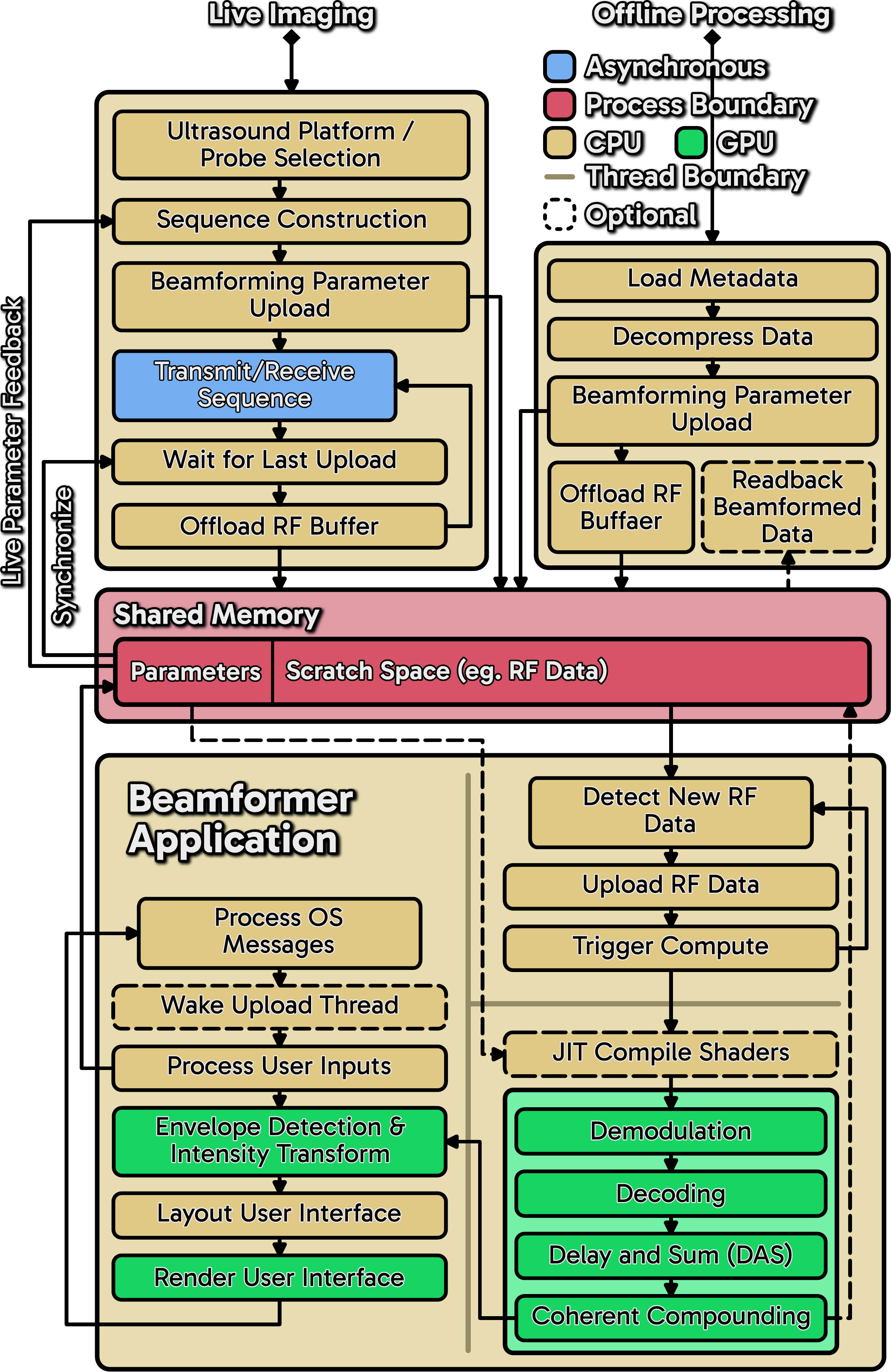}
	\caption{Detailed overview of the implemented software and external integration.
	         There are two primary ways to utilize the software. Shown on the left is
	         the live imaging method which makes use of a modified sequence sending data
	         to the beamformer. In this mode the beamformer can send modified parameters
	         back into the imaging sequence. The offline mode is similar; however, here
	         its common for the data to be read back and used for further analysis (SNR
	         measurements, resolution measurements, etc.).}
	\label{fig:software_details}
\end{figure}

The implemented beamformer supports both live processing,
utilizing communication with a connected imaging system, and
offline processing using a programming interface. From the
beamformer's perspective there is no difference between these two
modes. In Figure \ref{fig:software_details} we provide a detailed
overview of the control flow of the program in both cases. While
example implementations for the upper portion of the diagram,
which describes live and offline data processing, are available,
they are designed to be easily replaced by the user of the
software.

To process a data set the user must provide a description of the
data and a set of commands which should be used to process it.
The beamformer presents a typeless generic interface for passed in
data so the description is important to allow the data to be
processed correctly. It contains the data type, the method of
acquisition, the sample count, sampling rate, and so on. Next the
parameters related to beamforming are required. Two parameters
related to the array geometry must be passed: the element pitch
for the rows and columns, and a $4\times4$ affine transformation
matrix used to map from a global coordinate system (used to define
the origin for emitted waves) to one whose origin lies at the
center of the corner element of the receiver array and whose
$z$-axis is oriented with the normal of the array. In effect, we
treat the receiver array as a camera located in a 3D space. This
was done in order to support methods where multiple transducers
are utilized such as the case of future tiled arrays which are in
development. In the common case where a single array is used for
both transmission and receiving the transformation will simply
translate the origin to the center of the element at the array's
corner. Finally a set of parameters used for forming the image
should be provided. This includes the region and resolution of the
image, as well as information such as a mapping between data
channels and array channels, the orientation of the RCA on
transmit and receive, and the position of the transmit focal
point, which encodes the type of wave that was emitted
(cylindrical, planar, etc.). These parameters may be uploaded in
bulk using the simple API or in multiple parts with the advanced
API, useful for modifying parameters between runs or for utilizing
multiple parameter sets which allows the user to pre-upload a
batch of different options. This may be used for sequences
containing multiple acquisition variations interleaved together
such as the Cross Plane method shown.

\subsection{Metaprogramming}

There are many declarations, such as enumerations, flags, and
structure definitions, which are shared between the shader
programs, the C source code, and the library interfaces (e.g.
MATLAB). To avoid numerous mistakes introduced by having to
maintain many parallel files we implemented a metaprogram (a
program which can generate code) into our custom build process. It
parses a file containing definitions which should be shared and
outputs source code which is included later on in the build. For
example, the C portion of the output contains not only the
previously mentioned items, but also many tables mapping shader
IDs to shader source code; acquisition mode IDs to strings used
for displaying in the UI; mappings between C structure members,
used to pack the variables for baking into JIT compiled shaders,
and their identifiers which are used in the shader source code;
and more. This provides much more flexibility than what is
typically possible with just a macro pre-processor and C compiler.
Additionally this allows many operations to be table driven, which
reduces duplicate code, and is generally less prone to bugs. For
release versions of the program, the build process also embeds the
shader source code directly into the executable. This removes any
requirement on file loading at runtime which eliminates an entire
failure mode of the application.

\section{OPTIMIZATIONS}\label{sec:optimizations}

In this section we discuss some of the performance enhancements we
applied in the implemented beamformer. While many of them are
specific to the task being performed there is one global
consideration which we must always make when working with large
amounts of data. Both CPUs and GPUs have sets of cache memory
which is significantly faster to access than main memory (RAM or
VRAM) \cite{drepper2007every}. Cache is ordered in levels (L1, L2,
...) and lower levels are physically closer to core of the
processor and are therefore faster to access. Apart from atomic
operations, which may avoid the lowest cache levels, all memory
accesses will first try to find the data in the cache. If it is
not already in the cache the processor won't just read the bytes
that were requested, it will read a larger amount known as a cache
line, typically 64 bytes on a CPU, or on a GPU 32, 64, or 128
bytes depending which bytes are needed across the concurrently
running threads \cite{cuda2025}. Once the data is in the cache we
want to perform as many operations with it as possible before
requesting more from main memory. This is critical to ensure that
the processor is able to complete operations near its advertised
rate, particularly when dealing with large datasets, which we
define as any size of data which does not fit in a particular
level of the cache. We will refer to these caches further in the
following sections.

\subsection{Vantage Sequencing}\label{sec:vantage}

Transferring data from the Vantage system can be a major
bottleneck in the beamforming process. The first optimization is
to decouple the hardware acquisition rate from the software
processing speed. The Vantage system has two separate sequencers.
A hardware sequencer on the ultrasound platform and a software
sequencer on the host controller. These two sequencers can be
synchronized in various ways, and Verasonics' official
documentation details a couple of standard synchronization
methods; namely Serial, Synchronous, and Asynchronous acquisition.
The implemented beamformer can work with any of these modes, but
for an optimal live imaging experience we utilize Synchronous
mode. Unlike Serial acquisition, this removes any dependence on
the software processing from the hardware acquisition rate,
provided that the software processing rate is at least as fast as
the hardware acquisition rate, and unlike Asynchronous acquisition
this constrains the latency of the imaging system to be less than
the sum of the acquisition time, transfer time, and processing
time. To maximize the data throughput of Direct-Memory-Access
(DMA) transfers from the Vantage system to the host controller, we
have to optimize the transfer size as well. If the transfer size
is too low, then overhead time will dominate the transfer time,
and lower the overall transfer rate.  The transfer size has two
possible upper bounds: first it must fit within the on-system data
storage buffers, and next it must be small enough to completely
transfer to the host before the next acquisition completes.
Without satisfying the latter, the imaging system will be limited
by the transfer time and not by the acquisition time. This is
unavoidable if the maximum transfer rate is lower than the maximum
acquisition rate. For example, suppose we wish to image with a
5MHz array sampling at a rate of 20MHz to a depth of 10cm in a
media with a speed of sound of 1540m/s. We will acquire $\sim$2600
samples per receive channel per transmit. If we are performing
FORCES or HERCULES with 128 transmits and the samples are 16-bit
(2-bytes each) we will have 162.5MB/Frame. Assuming we can use the
full PCIe3.0 x8 (7.8GB/s) bandwidth on the Vantage's host
connection we will be able to acquire at most 49 frames per second
(FPS). In this case we would need to use a PRF of 6.3kHz excluding
dead time during the receive phase. In our case we must consider
two additional limiting factors, first the bias switching time
between transmits, and next a Vantage-to-host connection over
PCIe3.0 x4 as mentioned in Section \ref{sec:hardware}. For the
first, a limit in the current generation of biasing electronics
means that we must allow for a 250$\mu$s dead time between
triggering the biasing electronics and starting a transmit. This
limits us to a PRF of 4kHz resulting in an imaging rate of just 31
FORCES/HERCULES frames per second. Furthermore, the PCIe3.0 x4
connection to the Vantage means that we can only achieve a
transfer rate of 3.9GB/s. This limits us to a maximum of 24 FPS
(PRF of 3kHz) with our current hardware. In addition to the previous
points, as the Vantage system requires interfacing with MATLAB, we
are unable to avoid additional transfer time due to at least one
memory copying operation which must occur to transfer data from
the CPU to the GPU. This additional time shows up as added latency
in the image display. It does not affect the overall framerate
unless it is larger than the difference in computation time and
acquisition time. This overhead could potentially be avoided by
employing a similar strategy to \cite{tomov23ten}.

While we have discussed details of the Verasonics Vantage system
with the previous optimizations, they are applicable to other
ultrasound platforms. The implemented beamformer is hardware
agnostic and can be used with any ultrasound system given
appropriate software interfacing.

\subsection{Data Transfer}

\begin{figure}
	\includegraphics[width=\columnwidth]{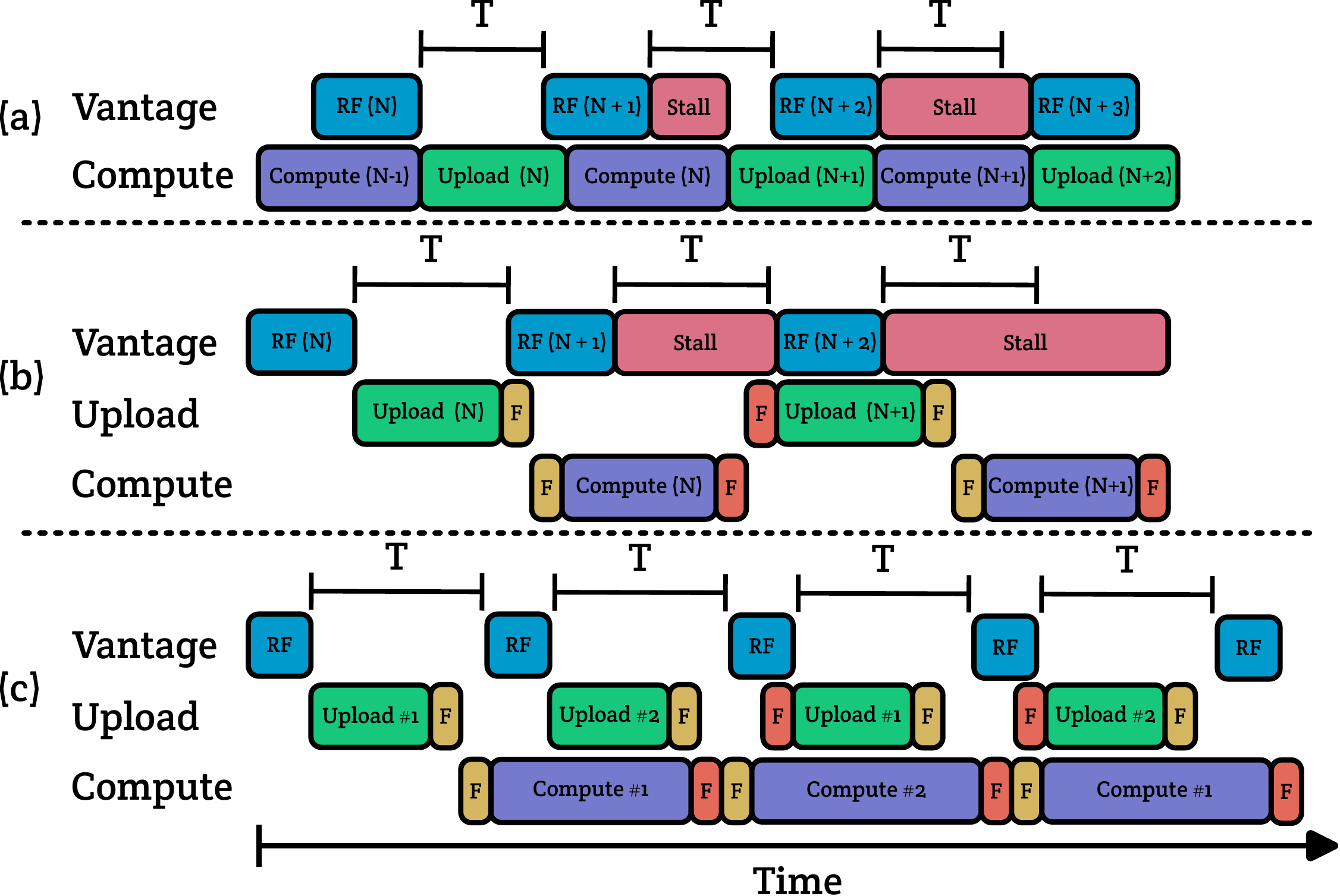}
	\caption{Different modes in which data handling could cause the Vantage acquisition
	         thread to stall. Here T represents the desired imaging period, which is not the
	         same as the PRF, and F represents a memory fence which provides a synchronization
	         barrier for the GPU. In (a) we utilize a single GPU hardware queue for both compute
	         and data upload. New RF data must wait for old RF data to be processed before it
	         can be uploaded. In (b) we make ineffective use of a separate hardware upload queue
	         which still introduces stalling. This method has only a single RF data buffer
	         available and cannot be overwritten until beamforming has finished. Finally in
	         (c) we solve the issue by allowing multiple RF data frames in flight.}
	\label{fig:data_stall}
\end{figure}

As we described in Section \ref{sec:hardware} our current imaging
setup is primarily limited by data bandwidth between our host
system and the Vantage system. However it was still important that
we eliminated any possible stalls caused by a data transfer to the
beamformer. The way which this occurs is demonstrated in Figure
\ref{fig:data_stall}. Suppose we wish to acquire with a constant
RF acquisition period of $T$. We must ensure that the sum of the
beamforming time and the data upload remains below $T$. If the
beamformer spends all of $T$ on beamforming the time needed to
upload the data would cause the beamformer to fall behind the
acquisition hardware which is shown in Figure
\ref{fig:data_stall}(a). Many modern GPUs have separate hardware
queues for computation and data transfer but access to them
requires special programming. While this is not directly possible
in OpenGL, most drivers will schedule data transfer on a dedicated
queue if a particular OpenGL context contains only data transfer
commands \cite{OpenGLInsights}. To achieve this we introduced a
CPU thread dedicated to uploading RF data to the GPU. On its own
this was insufficient to prevent sequence stalls due to data
uploading. As demonstrated in Figure \ref{fig:data_stall}(b), if
on a particular frame $N$, the combined GPU upload and compute
time took close to the target time $T$, a stall would occur when
uploading frame $N + 1$. This is because the compute of frame $N$
had not yet completed. To ensure this did not occur, we allowed
multiple frames to be in flight (in the process of beamforming) at
any time. This case is shown with 2 frames in flight in Figure
\ref{fig:data_stall}(c). In practice we extend this, allowing a
total of 3 frames in flight at any particular time. This is
facilitated by the use of a 3 frame GPU side ring buffer. Once the
upload of an RF frame into slot 1 has completed, a fence, a
synchronization primitive used to prevent operations in other
threads from proceeding until all threads have reached the same
point in the execution stream, is signaled indicating that the GPU
may start using the data. After the first stage of beamforming has
completed, a second fence is signaled indicating that slot 1 is
now free for a new dataset. The same sequence is repeated for
slots 2 and 3. Two sets of fences are required to ensure that
access to the buffer is valid. One set is used to prevent
partially uploaded data from being operated on; and a second
prevents partially processed data from being overwritten.
Furthermore, to maximize performance, we tell the graphics driver
that we do not want it to perform any form of synchronization on
the RF data buffer. This makes the use of fences critical and
failure to properly utilize them can lead to system crashes.

It is important to mention that we do not use a direct DMA
transfer from the imaging hardware to the GPU. While this adds
some amount of latency between the acquisition time and the time
the beamformed image is displayed, we can minimize its impact by
performing some light operations on the data while we have access
to it on the CPU. Since the data lives in CPU RAM, in order to
transfer the data to the GPU the CPU will have to touch every
sample. We can therefore perform a channel mapping while the data
is in flight and reduce both the amount of data the CPU has to
access and the amount of data that has to be uploaded over PCIe.
Assuming we only use a single CPU core to complete the mapping and
only half the channels actually contain data, as is common in
typical RCA imaging setups, we can reduce both the CPU memory copy
latency by half and the PCIe upload latency by half.


\subsection{Shader Dispatch}

\begin{table}
	\centering
	\caption{Compute Kernel Dispatch Layouts}
	\label{tab:compute_dispatch}
	\begin{tabular}{p{4.6em}cc}
	                     & \textbf{Layout \{X,Y,Z\}} & \textbf{Mapping}              \\
	Filter               & \{128, 1, 1\}             & \{Sample, Channel, Transmit\} \\
	Decode ($>$ 40 Tx)   & \{4,   1, 32\}            & \{Sample, Channel, Transmit\} \\
	Decode ($\le$ 40 Tx) & \{32,  1, 1\}             & \{Sample, Channel, Transmit\} \\
	DAS                  & \{16,  1, 16\}            & \{X, Y, Z\} Voxel             \\
	\end{tabular}
\end{table}

\begin{figure}
	\centering
	\includegraphics[width=0.8\columnwidth]{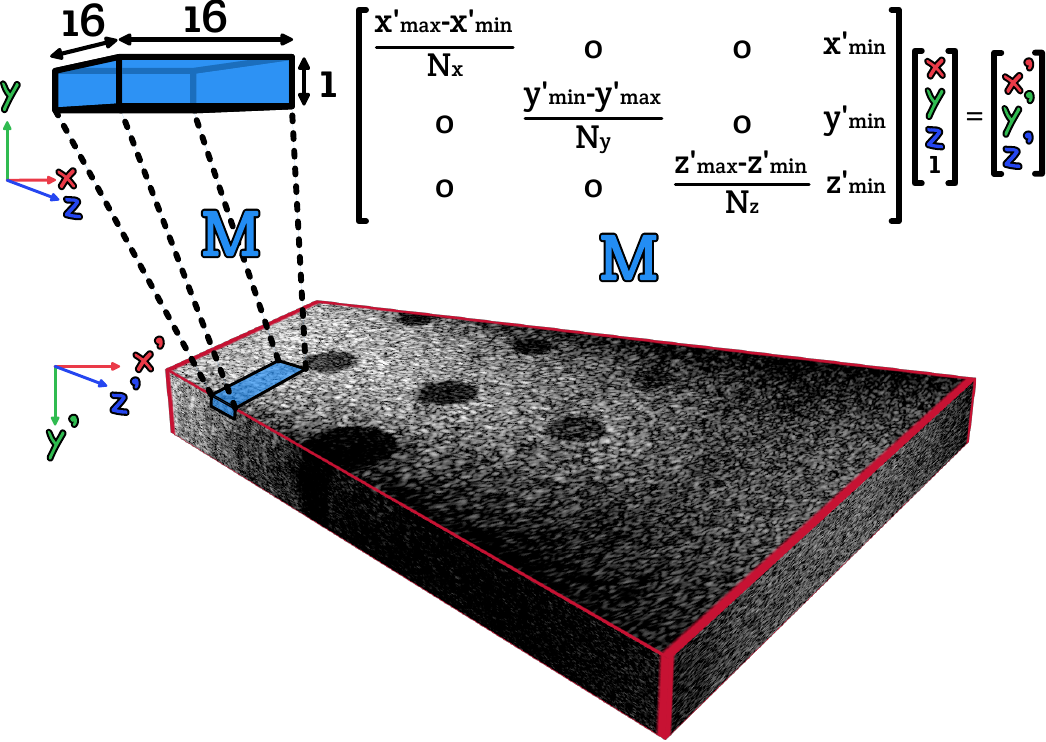}
	\caption{Mapping a dispatched thread group to an output region of a
	         beamformed volume. Here we use a matrix $M$ to transform a
	         single global thread index $\{x,y,z\}$ to a spatial point
	         $\{x',y',z'\}$ (e.g. in meters). $N_x,N_y,N_z$ represent
	         the number of image points and $x'_{\text{min},\text{max}}$,
           $y'_{\text{min},\text{max}}$,$z'_{\text{min},\text{max}}$ are
           the reconstruction region boundaries.}
	\label{fig:das_dispatch}
\end{figure}

There are two aspects of shader dispatch which we consider in
optimization of our pipeline. First we consider the overall
ordering of the individual processing stages. Then we provide
insight into how to dispatch each stage, i.e. the number of
threads dispatched in each dimension.

Most data transformations performed in the reconstruction process
are linear operations and can therefore be run in any order. While
this is mostly arbitrary there are few points to consider. First,
if we wish to perform decimation (which is a non-linear operation)
on the data it would be best to perform that as early as possible
to reduce the amount of data processed by any following stage.
Decimation can only be performed when the data is oversampled,
which typically only happens when we are utilizing quadrature
data. If the quadrature data must be demodulated, which involves a
low pass filtering step, that must occur prior to decimation.
Therefore we almost always perform demodulation as the first step.
Furthermore the filter coefficients are chosen to not introduce
any signal gain $> 1$ which allows us to convert back to 16-bit
integer data before storing the filtered result. This means that
the amount of data being processed will not expand for the next
stage and will decrease if we decimate.

The processing of FORCES and HERCULES datasets requires an
additional decoding step which we perform next. As will be
discussed in Section \ref{sec:optimizations}.\ref{sec:decoding},
this decoding step can be greatly accelerated if the data is first
reordered from the order provided by the imaging system. If we
don't run the decoding stage first we can configure the previous
stage to directly output data in the optimized layout and skip an
extra processing step.

The dispatch layout of individual shader stages is also important.
First, in order to saturate the GPU and reach a high thread
occupancy, we must ensure that the total number of threads in a
work group is not smaller than the number of hardware threads,
also known as the SIMD width, in a Compute Unit (CU). On NVIDIA's
GPUs this is 32 \cite{ada_arch}, and on AMD's GPUs this is 64,
though these are still split into 2 groups of 32 \cite{rdna4}.
While scheduling more threads than this is fine in most cases, as
they will run in sequence, scheduling less harms performance.
Similarly scheduling an amount more than this which is not an
integer multiple also harms performance as the GPU must still run
64 (32) threads at time and will never coalesce across groups.
Furthermore only 1024 threads are allowed to be scheduled at once
on most GPUs. Therefore we ensure all dispatch dimensions are at
least 64 threads wide, since we support both GPU vendors. The
actual distribution between the X, Y, and Z dispatch dimensions is
dependent on the shader. For demodulation/filtering, which only
operates on the time dimension, we dispatch 128 threads in X
allowing for optimal reuse of cached data samples. The decoding
shader also only cares about the 1D transmits dimension; however,
it must also sample from a 2D Hadamard texture. Texture sampling
on the GPU uses hardware optimized for multi-dimensional tiled
access \cite{tex_fetch}. To take advantage of this, for
$N_\text{tx} > 40$, we do a 2D dispatch with 4 threads in X, the
time sample/Hadamard column dimension, and 32 threads in Z, the
transmit/Hadamard row dimension. For $N_\text{tx} \le$ 40 we use a
different layout described in Section
\ref{sec:optimizations}.\ref{sec:decoding}. Finally, for DAS we
assign one thread to each output point (voxel or pixel). Because
data samples are similar within a small tiled region and the
typical case is 2‑D live‑view beamforming, we launch 16 threads
along the X (lateral) axis, 1 thread along Y (elevational), and 16
threads along Z (axial). We apply an affine transform (4$\times$4
matrix) to the desired output image plane such that it will always
map back to the X-Z thread dimension. The layouts of each kernel
what they map to is summarized in Table
\ref{tab:compute_dispatch}. A visual example of the mapping
between kernel dispatch dimensions and an output region of the
beamformed image is shown in Figure \ref{fig:das_dispatch}. It is
important to note that naive dispatch layouts can often cause
significant performance degradation, especially when large amounts
of memory are involved. As an example, the original dispatch
layout we tried for decoding was 32$\times$32$\times$1 with the
mapping time sample$\times$channel$\times$transmit. The extremely
poor memory access pattern in this case doubled the shader's
runtime over the current 4$\times$1$\times$16 layout.

\subsection{Filtering \& Demodulation}

\begin{figure}
	\centering
	\includegraphics[width=\columnwidth]{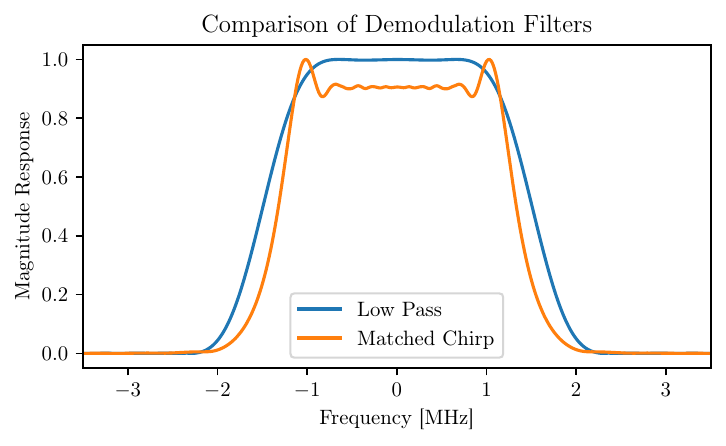}
	\caption{Frequency magnitude response for two different filter types used during the
	         demodulation stage of beamforming. For demodulation purposes we care about
	         the attenuation of high frequency components introduced by beating with the
	         carrier wave. Matched filtering also attenuates frequencies apart from the
	         those present in the matching waveform, however, it additionally applies
	         a nonlinear phase response which attempts to compress the pulse in time
	         restoring it to a delta.}
	\label{fig:filter_frequency_response}
\end{figure}

Most research ultrasound systems do not perform demodulation on
their own and instead provide data satisfying the sampling
requirements of the full RF carrier frequency. For bandlimited
signals, such as the ultrasound signal received by an ultrasound
probe \cite{gehlbach1981digital}, this signal is far oversampled.
To reduce the data processing requirements it is often beneficial
to demodulate the signal to baseband, decimate to a much reduced
sampling frequency, and perform any calculations on the complex IQ
signal. The process of demodulating a bandpass signal is well
documented in any signal processing textbook
\cite{RGLSignalProcessing}; however, we go a step further. To
obtain the maximum possible SNR from our receive signal it is
beneficial to apply a matched filter \cite{turin1960matched}.
Furthermore, if we apply a time domain encoding to the transmit
pulse, such as a chirp, a matched filter is critical to the
recovery of the desired signal \cite{misaridis2005modulated}. By
examining the frequency domain response of our desired matched
filters we observed that they also applied the low pass filtering
behaviour required for demodulation while applying the phase
correction necessary for signal recovery. Figure
\ref{fig:filter_frequency_response} demonstrates this effect on
the magnitude component of the signal's frequency components.
Therefore we need only apply the (baseband) matched filter during
demodulation to also gain the desired contrast enhancement.

We also applied an enhancement when demodulating. The typical
demodulation procedure involves a multiplication of all samples
with a complex exponential at the carrier/demodulation frequency,
assuming the IQ signal was sampled appropriately as in Section
\ref{sec:theory}.\ref{sec:sampling}. This is shown in Equation
(\ref{eq:demodulation}):

\begin{equation}\label{eq:demodulation}
	\text{IQ}[n] = \text{LPF}\left\{s[n]e^{-j\frac{2\pi f_c}{f_s}n}\right\}
\end{equation}

where LPF represents a low pass filtering operation, $f_c$
represents the carrier frequency, and $f_s$ represents the
sampling frequency. To evaluate this expression on the GPU we make
use of Euler's formula:

\begin{equation}
	e^{-j\frac{2\pi f_c}{f_s}n} = \cos\left[\frac{2\pi f_c}{f_s}n\right]
	                              - j\sin\left[\frac{2\pi f_c}{f_s}n\right]
\end{equation}

and perform a complex multiplication with the quadrature signal.
For the filtering operation to be applied (i.e. a convolution) to
a sample $n_i$, we require all samples in the range $(n_i - N_f,
n_i]$, where $N_f$ is the number of filtering coefficients. Since
the filtering of all time samples in the range $[n_i, n_i + N_f)$
require sample $n_i$, many duplicated loads and conversions will
be performed. We can avoid this by utilizing the GPU's Local Data
Share (LDS) functionality. LDS allows the programmer to
preallocate a portion of the GPU's L1 cache as a temporary buffer
which will be coherent within a single thread group. For a thread
group containing $N$ threads we may use LDS to cooperatively load
and convert $N + N_f$ samples and avoid most of the duplicated
work. This also allows us to efficiently pad the start of the
signal with 0s which helps to avoid thread divergence in the
convolution portion of the calculation. Applying this optimization
yielded a $\sim$30\% performance boost over the case where we
didn't share work.

\subsection{Decoding}\label{sec:decoding}

Both the FORCES and HERCULES methods make use of a spatial
Hadamard aperture encoding across a sequence of transmits
\cite{CC19_FORCES, dahunsi25hercules}. In order to utilize the
acquired data it must first be decoded by performing a matrix
multiplication across all transmits (see Section
\ref{sec:theory}.\ref{sec:sta}). Numerous optimized linear algebra
libraries exist for performing such a matrix multiplication.
cuBLAS \cite{cuBLASManual} is one such library provided by NVIDIA
for performing the operation on a GPU using CUDA. There are two
main limitations to this: first, the data must be in 32-bit
floating point format, but most imaging systems including our
Vantage-256, provide data as 16-bit integers. Therefore in order
to use the library we must first convert all data which doubles
its size. This greatly harms the performance since our data
pipeline is already heavily limited by memory throughput.
Secondly, we did not want the software to require an NVIDIA GPU to
be usable. Therefore we implemented our own matrix multiplication
routine in GLSL. Two optimizations were performed. First we
allowed it to operate directly on the smaller 16 bit data
directly. GLSL does not provide native support for 16-bit integers
without an extension so the data was accessed as if it was a
32-bit integer and unpacking was implemented manually. Since time
samples were stored next to each other in memory, and each time
sample can be decoded independently, we allowed each shader
invocation to decode two at a time. This gave a performance
improvement of $\sim$40\%. The second optimization we performed
was a first pass data reordering. To achieve better GPU L1 cache
utilization we desired for each data access to be sequential. Our
matrix multiplication operates across transmits so we need the
same time sample from different transmits to be stored next to
each other in memory. This is achieved by copying the data to a
second buffer which has the correct order. If we run a stage prior
to decoding, such as demodulation, we can tell that stage to store
the data in the correct order when its finished, but even when we
want to run the decoding stage first having an extra step which
only performs reordering still provides a significant performance
boost $\sim$10\%; the time spent performing a copy is amortized by
the speed-up produced by accessing sequential data elements.

Furthermore for transmit counts $>40$ we utilize local data
sharing to cooperatively preload all needed samples (columns of
the matrix) in parallel. Since the number of arithmetic operations
performed per sample in a matrix multiplication is low (1
fused-multiply-add (FMA)) it is beneficial to modify the algorithm
to operate on multiple rows of the second matrix, this is often
referred to as increasing the Arithmetic Intensity of the kernel.
In our current implementation 2 rows/thread gave the best
performance. The improvement varies by the number of transmits but
can be as high as 50\% for large transmit counts. For transmit
counts $\le 40$ this optimization will hinder performance since
the overhead added to perform synchronization outweighs any
caching inefficiencies. Instead we use a technique which we will
refer to as register caching. Here we preload all samples in an
input column into GPU registers and then calculate an entire
output column in a single thread. The higher register usage
prevents us from being able to fully saturate the GPU's work group
processor core and does not allow for maximum occupancy. However,
significantly higher arithmetic intensity combined with the fact
that the operation for each column index is independent from the
other indices allows the compiler to interleave ALU operations
which avoids the need to delay and wait for intermediate results.
With 40 transmits as an example this yielded a $\sim$33\%
performance boost.

\begin{figure}
	\includegraphics[width=\columnwidth]{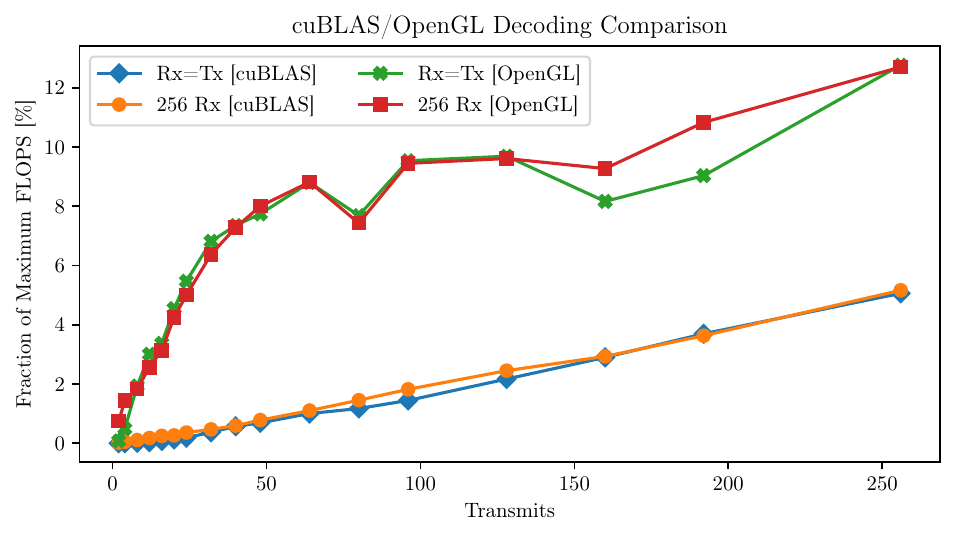}
	\caption{Comparison of RF Decoding performance between the GLSL implementation
	         and the cuBLAS implementation. cuBLAS makes use of highly optimized GeMM
	         algorithms which are designed for large, tightly packed matrices (matrices
	         which are stored sequentially in memory). Timing data shown here also includes
	         any time needed for converting (necessary for cuBLAS) and reshaping. We present
	         fraction of maximum device Floating Point Operations Per Second (FLOPS) as it
	         removes the dependency on a specific device.}
	\label{fig:decode_stats}
\end{figure}

We tested the performance of the implementation with 4096 time
samples for all supported Hadamard dimensions up until
256$\times$256. We include data reordering in this measurement
since we consider it part of the optimization even if it is
normally performed by the demodulation stage. The performance was
measured for both the case when the number of transmits is equal
to the number of receivers, as in full FORCES and HERCULES, and
the case when we transmit a sparse subset of the elements and
receive on all 256 channels, as is the case for uFORCES. This was
compared with a naive implementation using cuBLAS (using
\texttt{cublasSgemm}). The results are shown in Figure
\ref{fig:decode_stats} where we plot the fraction of total
available device Floating Point Operations Per Second (FLOPS), eg.
81.58 Tera-FLOPS on the RTX 4090, versus the number of transmits.
This was done to decouple the performance metric from a particular
GPU. The OpenGL trend follows identically on tested GPUs from AMD.
Our implementation provides significantly higher device
utilization then the naive cuBLAS implementation for all tested
Hadamard matrix sizes. There are two reasons for this, first the
memory traffic in the cuBLAS version is doubled due the need to
use 32-bit values. Note that even in the case of 16-bit values the
data is too large to fit in the L2 cache so there will be many
accesses that incur the full penalty for a VRAM access. Second,
the General Matrix Multiply (GeMM) algorithm employed by cuBLAS is
optimized for large matrices \cite{nugteren18clblast,
huang2020strassen} (one dimension $>$512). For the small matrices
we use here we do not benefit from the optimized GeMM algorithms.
Note that our implementation is not even close to fully utilizing
the GPU, we only reach 12.5\% utilization in the best case, but
this is still significantly higher than the 5\% peak utilization
achieved by cuBLAS.

\subsection{DAS}

Of all the shader stages utilized, Delay-and-Sum (DAS), as
described by Equation \ref{eq:das}, requires the most operations.
The total number of operations is further multiplied by the number
of output points or voxels the user has requested. DAS'
performance is made worse by the fact that there is no obvious way
to order the dispatch to share data between threads in a work
group. The dataset is large, larger even than the L2 cache on the
RTX4090 (72 MB), so we expect that many roundtrips to main memory
(VRAM) will be made. While it is not obvious how we could help the
GPU in preloading the data into cache we can do our best to ensure
that for any particular warp (group of threads) we only touch a
small subset of the data and thus allow it to remain in cache as
long as possible. We achieve this by having a warp beamform a
small tile/cube of the total B-Scan/volume. Within this region the
calculated times of flight will be similar meaning the data
samples that are needed will be closely grouped in memory. This
may allow for memory accesses across threads to be coalesced.
Furthermore we beamform with only a single channel of receive data
per dispatch. A single channel of receive data is usually small
($\sim$1 MB) which fits in the L2 cache. This can help to reduce
memory access latency for subsequent warps, should they utilize
the same portion of the receive data, since the L2 cache is
typically an order of magnitude faster to access than VRAM
\cite{mei2017dissecting}.

The best optimization that we can perform is to avoid reading the
receive data whenever possible. Since we are applying a dynamic
receive apodization we can skip sampling the data whenever the
apodization, calculated by Equations
(\ref{eq:receive_apodization}) and (\ref{eq:window}), is 0. This
occurs when the argument in Equation
(\ref{eq:receive_apodization}) is $\ge 0.5$. It is important to
test based on this value instead of evaluating the windowed value
since the evaluation of cosine will create a serialization point
and can take many cycles \cite{wong2010demystifying}. This method
is most effective for the FORCES, TPW, and VLS methods as their
receive element position is independent of the transmit index
allowing for the entire transmit batch to be skipped with just a
single apodization check.

\subsection{Compile Time Parameters}

The control flow of a GPU shader program is much more restrictive
than a program run on a general purpose CPU. This allows the
shader compiler to perform very heavy optimizations. We can take
advantage of this by providing the compiler with as much prior
knowledge as possible. This means that any value which an end user
may treat as variable; for example sample count, channel count, or
frequency; but is constant through a single run of the shader
program should be told to the compiler. This cannot be done if the
value is read out of a buffer of variables at runtime. Instead we
perform Just-In-Time (JIT) compilation of the shader program
whenever the user adjusts the configuration. This JIT process
involves preprocessing the shader source at runtime and inserting
the current value of each constant into the source code before
passing it to the compiler. Since the constants are not expected
to change very often and since shader compilers are designed to be
very fast (with this use case in mind), modifying variables does
not result in any user perceivable slowdown at runtime, even for
inputs coming from the UI. This flexibility is entirely afforded
to us by using shader programs rather than CUDA kernels. While it
would technically be possible to achieve with CUDA it requires
significant additional complexity: write the PTX (NVIDIA assembly
code) directly, which can be passed to the driver at runtime, or
load the CUDA code dynamically at runtime as a shared library,
detect that it needs updating, write the new source code to a
temporary file, launch an external process (nvcc), unload the
existing library, load the new library, and so on.

There are a number of benefits that compile time known constants
introduce. First, they avoid additional register usage by allowing
the compiler to bake constants directly into instructions as
immediate values, which allows for more warps to be scheduled at
the same time. This improves latency hiding since the GPU can only
schedule the number of warps supported by the available registers
at once. Additionally, constants can be folded (combined at
compile time), which avoids a number of extra arithmetic
instructions at runtime. Code branches based on constant values
can be eliminated at compile time. This includes those handling
loops in code, which can be unrolled. In the case of the decode
shader, which is simply performing a variable length dot product,
loop condition checking can be fully eliminated, which can reduce
the shader's runtime by $\sim$50\%.

\section{CONCLUSION}

In this work we have presented an open source GPU beamformer for
realtime reconstruction using conventional RCAs and TOBE arrays.
The software provides many programmable options for beamforming
and supports many imaging methods, some from our group such as
Fast Orthogonal Row Column Electronic Scanning (FORCES)
\cite{cc17_fast, MRS22_uFORCES} and Hadamard Encoded Row Column
Ultrasonic Expansive Scanning (HERCULES) \cite{dahunsi25hercules},
and some from other groups such as Virtual Line Source (VLS)
\cite{vls2002, rasmussen15three} and Tilted Plane Wave (TPW)
\cite{flesch17}. The software offers many options for
configuration. For example, in addition to the typical features
(region, resolution, etc.), features such as waveform matched
filtering, and the method of interpolation are also available.

Our implementation includes a number of features currently missing
from existing solutions. First, while a programming interface for
MATLAB is provided, the software can be used entirely
independently. This will make it possible to integrate with
systems driving it from other languages such as Python. Next we
implemented a 3D visualization for realtime cross plane imaging.
We propose that this view is better suited to navigation than
typical RCA methods owing to FORCES' ability to perform both
transmit and receive focusing in a B-Scan plane and image beyond
the shadow of the aperture. The software has both live imaging
capabilities when interfaced with an imaging system such as the
Verasonics Vantage, and offline reconstruction capabilities
through a programming interface. Finally the implemented software
is fully open source and released to the community under the
permissive Internet Systems Consortium (ISC) license.

In future works we will aim to further improve the beamforming
rate, either through new optimizations for the time domain delay
and sum, or through frequency-domain techniques such as F-k
Stolt's migration \cite{garcia13stolt} or others. Furthermore we
aim to introduce realtime processing for techniques such as Vector
Flow Imaging and ULM \cite{nahas2025bedside, ketterling2017high,
cohen2024multidirectional, zhou20183, wang2024ultrasound,
shin2025high, lowerison2025comparison, wu20253d} in addition to
new motion compensated Recursive Aperture Decoded Ultrasound
Imaging (READI) method \cite{henry2025recursive}. We also aim to
introduce new methods of realtime 3D volume visualization as well
as visualization for photo-acoustic imaging and visualizations
enabled by photo-acoustics such as lymphatic pumping
\cite{forbrich2017photoacoustic}.

\section*{CONFLICTS OF INTEREST}

RJZ and MRS are shareholders and directors of Clinisonix Inc. which
partially supported this work.

\section*{REFERENCES}
\printbibliography

\end{document}